\documentclass{article}
\usepackage{spconf,amsmath,graphicx}
\usepackage{graphicx}
\usepackage{textcomp}
\usepackage{xcolor}
\usepackage{algorithm}
\usepackage{algpseudocode}
\usepackage{amsmath}
\usepackage{setspace}
\usepackage{url}
\usepackage{breakurl}
\title{A knowledge-driven vowel-based approach of depression classification from speech using data augmentation}
\name{Kexin Feng and Theodora Chaspari \thanks{This work is supported by the National Science Foundation (CAREER: Enabling Trustworthy Speech Technologies for Mental Health Care: From Speech Anonymization to Fair Human-centered Machine Intelligence, \#2046118). The code is available at: \protect\url{https://github.com/HUBBS-Lab-TAMU/ICASSP-2023-Augmented-Knowledge-Driven-Speech-Based-Method-of-Depression-Detection}.}}
\address{Computer Science and Engineering \\
Texas A\&M University \\
\{kexin, chaspari\}@tamu.edu}
\begin{document}
%\ninept
%
\maketitle
\begin{abstract}
\vspace{-5pt}
We propose a novel explainable machine learning (ML) model that identifies depression from speech, by modeling the temporal dependencies across utterances and utilizing the spectrotemporal information at the vowel level. Our method first models the variable-length utterances at the local-level into a fixed-size vowel-based embedding using a convolutional neural network with a spatial pyramid pooling layer (``vowel CNN"). Following that, the depression is classified at the global-level from a group of vowel CNN embeddings that serve as the input of another 1D CNN (``depression CNN"). Different data augmentation methods are designed for both the training of vowel CNN and depression CNN. We investigate the performance of the proposed system at various temporal granularities when modeling short, medium, and long analysis windows, corresponding to 10, 21, and 42 utterances, respectively. The proposed method reaches comparable performance with previous state-of-the-art approaches and depicts explainable properties with respect to the depression outcome. The findings from this work may benefit clinicians by providing additional intuitions during joint human-ML decision-making tasks.
% , and also promotes the mitigation of gender bias.
\end{abstract}
\vspace{-10pt}
\section{Introduction}
\vspace{-10pt}
\label{sec: intro}
Depression is a mental health (MH) condition with large worldwide prevalence \cite{smith2014world}, whose diagnosis and treatment is challenging due lack of access to MH care resources and stigma \cite{katon2013depression}. Speech-based machine learning (ML) systems have shown promising results in identifying depression due to their ability to learn clinically-relevant acoustic patterns, such as monotonous pitch and reduced loudness~\cite{cohn2018multimodal}. In addition, these systems can potentially mitigate social stigma and increase accessibility to MH care resources, since they can run locally on users' smartphone devices. Various ML models including support vector machines (SVM), convolutional neural network (CNN), and long short-term memory (LSTM) have been explored for depression estimation \cite{valstar2016avec}. However, the majority of these methods are designed independently of MH clinicians, thus depicting challenges in transparency and explainability.

Interactions between humans and ML are evolving into collaborative relationships, where the two parties work together to achieve a set of common goals, especially when it comes to complex and highly subjective decision-making tasks, such as the ones pertaining to MH care. An explainable ML model of depression estimation would allow clinicians to gain insights into the ML logic and decision-making processes and contribute toward better calibrating their trust to the model output~\cite{saraswat2022explainable}. Previously proposed conceptual frameworks for building human-centered explainable ML suggest that users may be able to develop a mental model of the algorithm based on a collection of ``how explanations" that demonstrate how the model works based on multiple instances~\cite{lombrozo2009explanation}. In addition, it is important to provide both global explanations that describe holistically how the model works, and local explanations that demonstrate the relationship between inputs and outputs~\cite{mohseni2021multidisciplinary}.

Here, we design an explainable ML model for depression classification based on speech. We leverage knowledge from speech production indicating that depression can influence the motor control and consequently the formant frequencies and spectrotemporal variations at the vowel-level~\cite{atal1982new}. We propose a vowel-dependent CNN (\textit{vowel CNN}) with an spatial pyramid pooling (SPP) layer that learns the spectrotemporal information of short-term speech segments (i.e., 250ms) throughout the utterance. The depression is estimated from a group of vowel CNN embeddings using another 1D CNN (\textit{depression CNN}). The vowel CNN captures depression information at the local-level from parts of speech that are theoretically postulated to be most affected by the MH condition~\cite{atal1982new}. The SPP layer maps utterances of any size into a fix-size embedding that contributes to modeling explanations at the utterance-level, which can provide a global view of the depression outcome. In order to further improve the performance of the model and address challenges related to the small sample size, we use two novel data augmentation methods that are applied during the training of vowel CNN (i.e., addressing imbalance between vowel distributions) and depression CNN (i.e., alleviating imbalance between the health and depression classes). Results indicate that the proposed system is comparable to or better than various deep learning systems for classifying depression from speech \cite{ma2016depaudionet,lam2019context,vazquez2020automatic,chen2022speechformer,ravi2022fraug}, and further provides explanations about its decision-making process to the user. Ablation studies further demonstrate the effectiveness of considered augmentation methods.

% We address this challenge by designing an ML system utilizing the Spatial Pyramid Pooling (SPP) layer \cite{he2015spatial}. The SPP layer reserves the speech structure by mapping any-size utterances into a fix-size embedding. This reserves the structure of the speech at the utterance-level, and also avoids the silence/noises between utterances.

% thus making it possible to model the utterances. This helps to keep r
% The SPP layer converts any-size convolutional output into a fix-size embedding by adopting a multi-scale pooling scheme. 
% Our system trains a vowel CNN with an SPP layer that learns spectrotemporal information from 250ms segments. Then we use the vowel CNN to extract fixed-size embeddings from utterances, and access the probability of depression through the utterance-level embeddings using depression CNN.
% Novel data augmentation methods are applied during the training of both CNN models. 
% with better transparency and explainability.
% The results of our system are comparable to or better than the scores reported in multiple advanced ML systems.
% with the ability to access depression probability on different levels of granularity.

\vspace{-10pt}
\section{Previous work} \label{sec: previous}
\vspace{-10pt}
% Traditional ML models such as SVM are used with a 74-dimensional acoustic feature to identify the depression on the frame level \cite{valstar2016avec}.
End-to-end ML models can effectively detect one's MH condition from speech. Ma {\it et al.} designed an end-to-end system (DepAudioNet) that uses a 1-dimensional CNN to encode audio features and a LSTM network to model the encoded audio embeddings \cite{ma2016depaudionet}. Sardari \textit{et al.} utilized a convolutional autoencoder on the raw speech signal to identify depression \cite{sardari2022audio}. Romero \textit{et al.} introduced an ensemble learning approach by training 50 CNN models with different initializations to address the challenge of local optimal \cite{vazquez2020automatic}. Other recent methods, such as the SpeechFormer, use a hierarchical framework to modeling spectral variations within and across speech frames, phonemes, words, and utterances \cite{chen2022speechformer}.
% Liu \textit{et al.} investigated the influence of factors such as ensemble schemes and speaking styles for speech depression identification \cite{liu2017ensemble}. 

Data augmentation can contribute to effectively classifying depression with limited labelled data. As part of CNN-Augm, Lam \textit{et al.} augmented the speech transcripts based on the topic of the clinical interview via manually identifying the most common topics, and added speech samples related to each topic to form an augmented dataset \cite{lam2019context}. 
% This requires direct access to the transcripts, and merging multiple datasets that include topics could also be challenging. 
% Another recent method proposed by Ravi \textit{et al.} does not require any context information \cite{ravi2022fraug}. By using different frame-width and the frame-shift parameters, they can obtain data samples at the different level of time-frequency resolutions and thus augmenting the less frequent classes. 
Ravi \textit{et al.} used different frame-width and frame-shift parameters to obtain data samples at various time-frequency resolutions \cite{ravi2022fraug}. Other speech augmentation methods include feature perturbation \cite{park2019specaugment}, altering the raw speech signal \cite{ko2015audio,cui2015data}, or generating new data via a GAN-based structure \cite{jin2021adversarial}.

% Park \textit{et al.} designed a perturbation method that can be applied to features derived from speech (i.e., log mel spectrogram) \cite{park2019specaugment}. Such perturbation either performs a time wrapping, or masks random consecutive frequencies, or masks random consecutive time steps. While the majority of the study alter the time or frequency domain or modify the speed and volumn of the speech \cite{ko2015audio,cui2015data}, another recent speech data augmentation methods use GAN-based structure generate the dysarthric speech from the perturbed normal speech \cite{jin2021adversarial}.

The contributions of this work are as follows: (1) In contrast to the majority of deep learning models on depression estimation that are not explainable~\cite{ma2016depaudionet,sardari2022audio,chen2022speechformer}, we propose a local explanation of the ML decision via modeling speech patterns at the vowel-level; (2) We extend the limited prior work on explainable ML for depression classification~\cite{feng2022toward} by further providing a global explanation of the decision at the utterance-level via introducing the SPP layer that can model utterances of variable length; and (3) We investigate an oversampling-based and perturbation-based augmentation methods to mitigate the imbalanced distributions of the different vowels and healthy/depression classes.

% Inspired by the observation that depression is often combined with a reduced vowel space \cite{scherer2015reduced}, Feng \& Chaspari assigned vowel labels to 250ms audio segments based on an aligner, and classify the segments using a CNN to model the spectrotemporal variations in speech vowels, which feature embeddings are used to identify the depression.

% In contrast to the majority of prior work in deep learning, this paper proposes a CNN-based architecture that learns vowel-based embeddings, therefore explicitly incorporating low-level spectrotemporal information that is valuable for identifying depression \cite{shimizu2005chaos, scherer2015self, scherer2015reduced, vlasenko2017implementing, stasak2019investigation}. While phoneme-based information has been modeled before \cite{muzammel2020audvowelconsnet}, it was only incorporated at the extraction of spectrogram features, without further deriving high-level representations;

% In this work, we introduce a method that models speech utterances with variable lengths but still maintains a promising performance on depression identification. 

\vspace{-10pt}
\section{Proposed Methodology} \label{sec: method}
\vspace{-10pt}
% The proposed system improves the previous research proposed by Feng and Chaspari \cite{feng2022toward} by cooperating designed data augmentation methods and a redesigned depression classification module for better explanation purposes. The added augmentation procedures can mitigate the challenges resulting from the imbalance between positive and negative samples that exist in commonly used depression datasets, and the uneven distribution between vowels and non-vowel segments. There are four modules in our designed system: (1) augmented vowel segmentation module (Section \ref{ssec: 1}), which samples almost evenly distributed 250ms segments for six vowel labels same as the previous research (/a/, /e/, /i/, /o/, /u/, or not a vowel); (2) vowel classification module (Section \ref{ssec: 2}), that includes a classification 2D CNN trained based on the balanced vowel segments and able to take variable-sized input; and (3) augmented depression classification module (Section \ref{ssec: 3}), that trains another 1D CNN to identify the depression with the augmented the vowel-based embeddings extracted from the previous step.
Our proposed system includes four modules: (1) Vowel segmentation module with data augmentation (Section \ref{ssec: 1}), which evenly samples 250ms segments from six English vowels (/a/, /e/, /i/, /o/, /u/, or not a vowel); (2) Vowel classification module (Section \ref{ssec: 2}), that includes a vowel CNN trained based on the balanced vowel segments. This vowel CNN can take variable-length utterances as input due to the SPP layer; and (3) Depression classification module with data augmentation (Section \ref{ssec: 3}), that trains the depression CNN using the vowel CNN embeddings.
\vspace{-10pt}
\subsection{Vowel segmentation module with data augmentation} \label{ssec: 1}
\vspace{-5pt}
% In the previous research by Feng and Chaspari, they extracted the 250ms segments with a 125ms overlap between each segment \cite{feng2022toward}. Due to the nature of the speech, the distribution of the vowels is highly imbalanced (i.e., /a/, the most occupied vowel, is over 10 times more than rare vowels such as /u/). Randomly oversampling the less occupied vowels mitigates this problem. Another augmentation method, as proposed in this work, is to dynamically determine the overlap between segments. More specifically, we keep a normal overlap (125ms) around a segment that is not a vowel, and a smaller overlap (i.e., 50ms) around a segment that is labeled as a vowel. Compared with randomly oversampling, it provides more varieties even on the least occupied vowels. A more detailed setting and specific examples for each vowel are included in Table \ref{tab: vowel_aug}. The ratio for each vowel is predetermined based on the training data. The other part of this module (i.e., assign a label for a given segment) is kept the same as the previous research \cite{feng2022toward}.  
In this module, we prepare the training data (i.e., 250ms speech segments with vowel labels) for the vowel CNN, similar to \cite{feng2022toward}. 
% Feng \& Chaspari also extracted 250ms segments from the training data, 
% but the sample distribution of vowel labels is highly imbalanced (i.e., /a/, the most occupied vowel, is over 10 times more than rare vowels such as /u/), which is due to the nature of the speech \cite{feng2022toward}.
The most common vowel (/a/) occurs over 10 times more than less frequent vowels (/u/) due to the linguistic of the English language, which can potentially hamper the vowel classification module from effectively learning vowel-dependent patterns.
Thus, we design a sampling-based data augmentation method by dynamically determining the overlap between segments. If the current segment (0-250ms) belongs to 'not a vowel', a normal length of overlap ($250ms*0.5=125ms$) will be applied (i.e., the next segment starts at 125ms). However, if the current segment is labeled as the vowel /a/, the overlap length will be reduced to 75ms ($250ms*0.3=75ms$) so that more segments can be sampled around this vowel. The overlap ratio for each vowel is predetermined based on the vowel distribution of the training data and remains unchanged in the sampling process. An example of this approach is provided in Table \ref{tab: vowel_aug}.
% Compared with random oversampling, this method provides increased variability in the generated data even for the less frequent vowel classes. 
A segment is finally assigned to a vowel based on whether the vowel is fully or partially included within this segment, similar to \cite{feng2022toward}.

\begin{table*}[t]
\caption{An example showing how the vowel segmentation with augmentation method determines the segment timestamp.}
\scalebox{0.9}{
\begin{tabular*}{1.1\textwidth}{@{\extracolsep{\fill}}ccccccc}
\hline
current segment start/end (ms) & \multicolumn{6}{c}{0-250}                           \\ \hline
current vowel label  & /a/   & /e/   & /i/   & /o/   & /u/   & not a vowel \\
overlap ratio (between current and next segment)       & 0.3   & 0.08   & 0.1   & 0.03   & 0.02   & 0.5         \\
next segment start/end (ms)    & 75-325 & 20-270 & 25-275 & 7.5-257.5 & 5-255 & 125-375       \\ \hline
\end{tabular*}}
\label{tab: vowel_aug}
\vspace{-10pt}
\end{table*}

\vspace{-10pt}
\subsection{Vowel classification module} \label{ssec: 2}
\vspace{-5pt}
The purpose of this module is to train the vowel CNN that classifies the extracted 250ms segments into their assigned labels. Compared with a vanilla 2D CNN used in previous work~\cite{feng2022toward}, the SPP layer increases the model's flexibility by allowing different embedding shapes after the convolutional layers.
% Using the augmented training set as in Section \ref{ssec: 1}, our designed model obtains more separable embeddings based on the vowel classification result. 
We use the log-Mel spectrogram as the feature for every 250ms segment. The feature extraction process is completed using the Librosa library \cite{mcfee2015librosa} and the other parameters include a 512-sample FFT window length, 128-sample hop length, and 128 Mel bands. This leads to a spectrogram patch with size $(128, 28)$ for every segment. The vowel CNN includes three convolutional blocks, each block consists of a convolutional, activation, batch normalization, and max-pooling layer. A SPP layer and two fully connected layers are further added after all convolutional blocks. We show the detailed vowel CNN structure in Table \ref{tab: vowel_cnn}. We use Pytorch \cite{paszke2019pytorch} to implement this model and minimize the cross-entropy loss. A batch size 64 and an Adam optimizer with a learning rate of 0.001 and l2 regularization of 0.001 are also applied.

\begin{table*}[t]
\caption{The structure and hyper-parameters of the 2D CNN model that conducts vowel classification.}
\scalebox{0.9}{
\begin{tabular*}{1.1\textwidth}{@{\extracolsep{\fill}}ccccccc}
% \begin{tabular}{ccccccc}
\hline
Layer & Conv block 1 & Conv block 2 & Conv block 3 & SPP & Fully-connected (FC) & Output \\ \hline
{Layer setting} & \begin{tabular}[c]{@{}c@{}}conv kernel (3, 1)\\ 64 filters, ReLU\\ pooling kernel (2, 1)\end{tabular} & \begin{tabular}[c]{@{}c@{}}conv kernel (3, 1)\\ 64 filters, ReLU\\ pooling kernel (2, 1)\end{tabular} & \begin{tabular}[c]{@{}c@{}}conv kernel (3, 1)\\ 64 filters, ReLU\\ pooling kernel (2, 1)\end{tabular} & 3 levels &  \begin{tabular}[c]{@{}c@{}}128 units\\ ReLU\end{tabular} & 6 units \\
{output dim} & (64, 63, 28) & (64, 30, 28) & (64, 14, 28) & 1344 & 128 & 6 \\ \hline
\end{tabular*}}
\label{tab: vowel_cnn}
\vspace{-12pt}
\end{table*}
% \vspace{-5pt}

\vspace{-10pt}
\subsection{Depression classification module with augmentation} \label{ssec: 3}
\vspace{-5pt}
This module utilizes the fixed-size embeddings at the utterance level extracted from the vowel CNN and classifies depression for each speaker using soft voting. Because the vowel CNN can take an arbitrary-sized input, we extract a 128-dimensional vowel embedding by taking the output of the first fully connected layer for each utterance. Based on these 128-dimensional embeddings, we train a depression CNN model, which takes a group of embeddings (i.e., 42 utterances) as input and outputs the prediction probability for this group. 
% Conducting decisions on utterances groups may provide better insights for each part of the speech compared with the LSTM model used in previous research \cite{feng2022toward}. 
Before training the depression CNN, we generate an augmented training set using a novel data augmentation method. Inspired by KeepAugment \cite{gong2021keepaugment}, our method randomly selects successive utterance windows from speakers, and performs perturbation without changing the utterances with high saliency measures, thus protecting the utterance with rich spectrotemporal information. The saliency measure for every utterance is obtained using the trained vowel CNN, and the perturbation is done by replacing this utterance with a fixed constant $\mathbf{c}$.
% value $\mathbf{v}$ that is either a fixed value of 0.001, or the mean value of the perturbed utterance embeddings.
% We explore two types of perturbing value $\mathbf{v}$: (1) a fixed value of 0.001; and (2) the mean value of the perturbed utterance embeddings. 
A detailed description of applying this data augmentation procedure to a speaker can be found in Algorithm \ref{Alg: aug}. The parameters for the augmentation include $\mathbf{n} =42$ (number of utterances included as an input in the depression CNN), $\mathbf{pos} =8$ (number of samples added to the training set if the original sample has depression label), $\mathbf{neg} =4$ (number of samples added to the training set if the original sample has a non-depression label), $\mathbf{p} =6$ (number of utterances that are perturbed for a group of utterances), $\mathbf{r} =21$ (number of utterances that are protected due to high saliency values), and $\mathbf{c}=0.001$ (perturbed value).

The augmented dataset consists of $(\mathbf{n}, 128)$-dimensional samples, where $\mathbf{n}$ is the utterance window size (empirically set to 42) and 128 is the dimension of the vowel embeddings. The structure of the depression CNN is in Table \ref{tab: depression_cnn}.
% We train a 1D CNN model with the structure shown in Table \ref{tab: depression_cnn} on this new dataset. 
This model minimizes the cross-entropy loss, trained using a batch size of 16, and an Adam optimizer with a learning rate of 0.001 and l2 regularization of 0.01. For a test speaker, we segment the speech into windows of $\mathbf{n}$ utterances without overlap and perform soft voting over all windows. We evaluate different utterance window sizes ($\mathbf{n}=21,10$), for which we also increase the $(\mathbf{pos}, \mathbf{neg})$ to $(16, 8)$ and $(32, 16)$, and the number of perturbed utterances $\mathbf{p}$ to $2$ and $1$, respectively. %The model structure and number of epochs are revised accordingly, but the other training parameters remain the same.% The model structures for different $\mathbf{n}$ are found in GitHub.

% We also explore the possibility of using a smaller utterance window size during both augmented training and testing ($\mathbf{n}=21$ and $\mathbf{n}=10$). In this case, we increase the $(\mathbf{pos}, \mathbf{neg})$ to $(16, 8)$ and $(32, 16)$, respectively, to fully utilize the available data. The number of perturbed utterances $\mathbf{p})$ is also proportionally reduced to $3$ and $1$. The model structure and epoch are also revised to fit the smaller input size, and the other training parameters are kept the same. The model structures for different $\mathbf{n}$ is included in the GitHub link due to the page limitation.
% \vspace{-5pt}
\begin{algorithm}[!htb]
\caption{Data augmentation in the augmented depression classification module}
\begin{algorithmic}
\State \textbf{Input:} 
\State --- vowel CNN $f$, utterance embeddings and depression label for a speaker $(\mathbf{X}, \mathbf{y})$, window size $\mathbf{n}$, \# augmented positive (negative) samples $\mathbf{pos}\ (\mathbf{neg})$; \# utterances perturbed $\mathbf{p}$; \# utterances protected $\mathbf{r}$; a constant $\mathbf{c}$
% ; perturb value $\mathbf{v}$
\State \textbf{Output:} set $Z$ of the augmented data for this speaker
\State \# Initialize the augmented set
\State $Z \gets \{\}$ 
% \Comment{Initialize the augmented set}
\State \# Different number of augmented samples based on label
\State $aug\_num = \mathbf{pos}$ if $\mathbf{y}=1$ else $\mathbf{neg}$ 
% \Comment{Generate different number of samples for each class}
\For{i from 0 to $aug\_num$} 
    \State Randomly get $\mathbf{n}$ sequential utterance embeddings in $\mathbf{X}$
    \State Obtain the saliency measure using $f$ for each utterance
    \State Draw $\mathbf{p}$ utterances not in top $\mathbf{r}$ saliency measures
    % \For{every utterance in selected $\mathbf{p}$ utterances}
    %     \State Replace this utterance with perturbed value $\mathbf{v}$
    % \EndFor
    % \State Replace select $\mathbf{p}$ utterances with perturbed value $\mathbf{v}$
    \State Replace this $\mathbf{p}$ utterances with $\mathbf{c}$
    \State $Z \gets Z \cup \{$perturbed utterance embeddings$, \mathbf{y})\}$
\EndFor
\State \Return{$Z$} 
\end{algorithmic}
\label{Alg: aug}
\end{algorithm}

\begin{table}[!tb]
\caption{The structure and hyper-parameters of final classification CNN in augmented depression classification module.}
\scalebox{0.9}{
\begin{tabular*}{1.1\linewidth}{@{\extracolsep{\fill}}ccccc}
\hline
Layer & Conv block 1 & Conv block 2 & FC & Output \\ \hline
{layer Setting} & \begin{tabular}[c]{@{}c@{}}kernel 7\\ 32 filters \\ ReLU\\ pool size 2\end{tabular} & \begin{tabular}[c]{@{}c@{}}kernel size 7\\ 32 filters\\ ReLU\\ pool size 2\end{tabular} & \begin{tabular}[c]{@{}c@{}}64 units\\ ReLU\end{tabular} & 2 units \\
{Output dim} & (32, 18) & (32, 6) & 64 & 2 \\ \hline
\end{tabular*}}
\label{tab: depression_cnn}
\vspace{-5pt}
\end{table}
\vspace{-5pt}
\subsection{Evaluation} \label{ssec: 4}
\vspace{-5pt}
We first measure the vowel classification performance of the vowel CNN (Section \ref{ssec: 2}). We also report the performance of the depression CNN (Section \ref{ssec: 3}). Finally, we explore the acoustic descriptors that correlate with the system's output. For every depression CNN's $\mathbf{n}$ utterance input, we extract 7 interpretable features that are related to depression: (1) speech percent; (2) mean value of the fundamental frequency (mean F0); (3) standard deviation of the fundamental frequency (std F0); (4) mean value of the jitter; (5) mean value of the shimmer; (6) mean value of the loudness. 
% The first three measures are obtained using the Librosa Python library \cite{mcfee2015librosa}, and we extract the remaining measures using the openSMILE toolkit \cite{eyben2010opensmile}.
These acoustic measures are obtained using the Librosa Python library and the openSMILE toolkit \cite{mcfee2015librosa, eyben2010opensmile}. We can quantify the relatedness of each acoustic measure with the model prediction, by calculating their Pearson's correlations.

\vspace{-20pt}
\section{Experiments} \label{sec: experiment}
\vspace{-5pt}
\subsection{Data description} \label{ssec: data description}
\vspace{-5pt}
We use the Wizard-of-Oz part of the Distress Analysis Interview Corpus (DAIC-WoZ) \cite{gratch2014distress} that includes 142 clinical interviews split into a training set (30 depressed, 77 non-depressed), a development set (12 depressed, 23 non-depressed), and a test set (labels are unknown to the public per the AVEC 2016 challenge) \cite{valstar2016avec}. A participant is considered to have depression if they score equal to or larger than 10 in the Patient Health Questionnaire (PHQ-8) \cite{gilbody2007screening}. We report the performance on the development set, which is consistent with previous research.

\vspace{-10pt}
\subsection{Baseline methods} \label{ssec: data description}
\vspace{-5pt}
We use two baseline methods that have considered data augmentation for depression estimation; the CNN-Augm \cite{lam2019context}, which requires content information, and the Ensemble-CNN \textit{et al.} \cite{vazquez2020automatic}, which includes 50 CNN models rendering it challenging to run locally on wearable or portable devices and extremely hard to explain the decision. We also report results from DepAudioNet \cite{ma2016depaudionet} and SpeechFormer \cite{chen2022speechformer} as additional baselines. All the above are discussed in Section~\ref{sec: previous}.
\vspace{-10pt}
\subsection{Results} \label{ssec: results}
\vspace{-5pt}
We first show the effect of data augmentation in the vowel segmentation module (Section \ref{ssec: 1}). We report the F1-score of vowel classification on the development set using random oversampling and our proposed vowel augmentation method in Table \ref{tab: vowel_result}. Our vowel augmentation method leads to better performance in every class. We identify significant differences in performance between models via a McNemar test ($p<0.001$).% a contingency table based on whether each model's prediction is consistent with the true class, and then perform a McNemar test ($p<0.001$) to identify significance between models.
% A \textcolor{red}{XX} significant analysis is performed between the models, and this improvement is significant with a $p<0.001$.
% But such improvement is less obvious compared with the vowel CNN trained in the previous research \cite{feng2022toward}. A possible reason could be the added SPP layer resulted in some information loss, even though it preserves more information compared with global max-pooling. 

\begin{table}[!tb]
% \caption{F1-score of vowel CNN trained with random oversampling and augmented vowel data.}
\caption{F1-score of vowel CNN.}
\scalebox{0.8}{
% \begin{tabular*}{\linewidth}{@{\extracolsep{\fill}}ccccccc}
\begin{tabular}{cccccccc}
\hline
 & /a/ & /e/ & /i/ & /o/ & /u/ & \begin{tabular}[c]{@{}c@{}}Not \\ vowel\end{tabular} & \begin{tabular}[c]{@{}c@{}}Macro \\ F1\end{tabular} \\ \hline
\begin{tabular}[c]{@{}c@{}}random \\ oversampling\end{tabular} & 0.64 & 0.4 & 0.5 & 0.34 & 0.33 & 0.66 & 0.48 \\
\begin{tabular}[c]{@{}c@{}}augmented \\ vowel CNN\end{tabular} & 0.66 & 0.41 & 0.51 & 0.36 & 0.35 & 0.69 & 0.50\\ \hline
\end{tabular}}
\label{tab: vowel_result}
\vspace{-10pt}
\end{table}

Then we present the F1-score on the depression class, and the macro F1-score obtained using the model in the augmented depression classification module (Section \ref{ssec: 3}) in Table \ref{tab: depression_result}. The proposed method provides competitive performance with different utterance window size $\mathbf{n}$, which adds extra flexibility in decision granularity (i.e., fine-grain decisions for $\mathbf{n}=10$; and coarse-grain for $\mathbf{n}=42$). 
% Our multiple experimental settings provide a higher precision compared with the previous research. 
We verify the effectiveness of the data augmentation method by removing the perturbation (i.e., set $\mathbf{p}=0$), and the performance declined for all the experimental settings ($\mathbf{n}=\{42, 21, 10\}$). Under a setting of $\mathbf{n}=42$ and $\mathbf{p}=0$, we obtain a macro-F1 at 0.64. The obtained score is slightly better than DepAudioNet, which applies a sampling-based augmentation without any perturbation. 

\begin{table}[!tb]
\caption{Performance on the dev set for different methods.}
\scalebox{0.85}{
% \begin{tabular*}{\linewidth}{@{\extracolsep{\fill}}ccccc}
\begin{tabular}{ccccc}
\hline
% Method & Precision & Recall & F1 & \begin{tabular}[c]{@{}c@{}}Macro \\ F1\end{tabular} \\ \hline
Method & Precision & Recall & F1 & Macro F1 \\ \hline
DepAudioNet (2016) \cite{ma2016depaudionet} & 0.35 & 1.0 & 0.52 & 0.61 \\
CNN-Aug (2019) \cite{lam2019context} & 0.78 & 0.58 & 0.67 & - \\
Ensemble-CNN (2020) \cite{vazquez2020automatic} & 0.55 & 0.79 & 0.65 & 0.73 \\
SpeechFormer (2022) \cite{chen2022speechformer} & - & - & - & 0.69 \\
Fraug (2022) \cite{ravi2022fraug} & - & - & - & 0.66 \\
Proposed ($\mathbf{n}=10$) & 0.55 & 0.50 & 0.52 & 0.64 \\
Proposed ($\mathbf{n}=21$) & \textbf{0.80} & 0.33 & 0.47 & 0.65 \\
% Ours ($\mathbf{n}=42$, mean-based $\mathbf{v}$) & \textbf{0.83} & 0.42 & 0.56 & 0.70 \\
% Ours ($\mathbf{n}=42$, fixed $\mathbf{v}$) & 0.70 & 0.58 & 0.64 & \textbf{0.73} \\ \hline
Proposed ($\mathbf{n}=42$) & 0.70 & 0.58 & 0.64 & \textbf{0.73} \\ \hline
\end{tabular}}
\label{tab: depression_result}
\vspace{-10pt}
\end{table}

Finally, we explore what acoustic attributes relate to our system when conducting a depression identification as described in Section \ref{ssec: 4}. We report the correlations between measures and model output of depression probability in Table \ref{tab: correlation}. %We find that the model output is significantly correlated with some acoustic attributes when using a small number of utterances, but such influence can be mitigated when increasing the value of $\mathbf{n}$. This indicates that including more utterances at the input of the depression CNN may contribute to a more informed decision based on different types of acoustic measures. 
Jitter and speech percentage are significantly associated with the model's depression likelihood, which might indicate that the proposed explainable model captures perceptually intuitive information about depression.
Mean F0 and loudness are the two most significantly correlated acoustic features with the model output for $n=10$. Based on prior work, these features are not always correlated with depression scores \cite{taguchi2018major,zhang2019evaluating}, but are highly indicative of a speaker's demography, and particularly gender~\cite{lee2010role}. This may indicate that the model captures gender information, as a result of the gender imbalance and higher prevalence of female speakers with depression (14 out of 63 male speakers with depression; 16 out of the 45 female speakers). While evidence of gender bias in this data also exists from prior work \cite{bailey2021gender}, a more detailed analysis is needed to fully understand this observation.

\begin{table}[!tb]
\caption{The correlation between acoustic measures and depression probability from the model with different values of $\mathbf{n}$. Bold indicates significant correlation ($p < 0.05$)}.
\scalebox{0.85}{
\begin{tabular}{ccccccc}
\hline
$\mathbf{n}$ & \begin{tabular}[c]{@{}c@{}}speech \\ percentage\end{tabular} & \begin{tabular}[c]{@{}c@{}}mean \\ F0\end{tabular} & \begin{tabular}[c]{@{}c@{}}std \\ F0\end{tabular} & \begin{tabular}[c]{@{}c@{}}jitter \end{tabular} & \begin{tabular}[c]{@{}c@{}}shimmer\end{tabular} & loudness \\ \hline
10 & \textbf{-0.119} & \textbf{0.396} & 0.043 & \textbf{0.13} & \textbf{-0.088} & \textbf{-0.324} \\
21 & \textbf{-0.195} & \textbf{0.377} & 0.088 & 0.15 & \textbf{-0.138} & \textbf{-0.411} \\
42 & -0.122 & \textbf{0.236} & 0.151 & 0.113 & -0.156 & \textbf{-0.213} \\ \hline
\end{tabular}}
\label{tab: correlation}
\vspace{-10pt}
\end{table}

% \vspace{-10pt}
% \section{Discussion} \label{sec: discussion}
% \vspace{-5pt}
% When $\mathbf{n}=10$, the two acoustic features significantly correlated with the model output are also closely related to gender differences.
% A model that relies a lot on these two attributes could identify a female as depressed, even if she says the exact same content as a male speaker that predicted health. Previous research also reported gender bias by observing the performance difference between male and female speakers \cite{bailey2021gender}. Such bias may result from the gender and class imbalance within the training data (63 males of which only 14 are depressed, and 45 females but 16 are depressed). It could also because when decreasing the $\mathbf{n}$ to $10$, less information is included in utterance windows, and thus the model has to rely more on 'easy' attributes to reach a local optimal. Mitigating the possible gender bias will become part of our future work.
\vspace{-5pt}
\section{Conclusions}
\vspace{-5pt}
We explored an explainable ML that integrated vowel-based information at the local level and modeled speech utterances with variable lengths with an SPP layer. A dynamic sampling-based data augmentation method was designed to address the distribution difference among vowels.
% We then train a 1D CNN using the embeddings extracted using vowel CNN from a group of utterances. 
We then trained a depression CNN that conducted the final classification utilizing the vowel CNN embeddings.
% To address the performance decline resulting from the SPP layer, we use a dynamic overlap to sample more segments around the less frequent vowels to benefit the training of vowel CNN. We also design a perturb-based data augmentation method to boost depression identification performance. 
A perturbation-based data augmentation was applied to further mitigate the class imbalance.
Our proposed system had a better or comparable performance compared to multiple baselines, along with increased explainability. As part of our future work, we plan to evaluate the proposed approach via user studies with MH experts and consider confounding factors such as gender to better disentangle MH information from other confounding factors.
% address the observed gender bias, and thus further improve the explainability of our system.

% \setstretch{0.9}
\renewcommand{\normalsize}{\fontsize{8.3}{10.70}\selectfont}
\normalsize
\vfill\pagebreak
\let\oldbibliography\thebibliography
\renewcommand{\thebibliography}[1]{%
  \oldbibliography{#1}%
  \setlength{\itemsep}{-5pt}%
}

\bibliographystyle{IEEEbib}
\bibliography{refs}

\end{document}